\documentclass{svjour2}
\usepackage{graphicx}

\def\beq{\begin{equation}}
\def\eeq{\end{equation}}
\def\rmd{{\rm d}}

\journalname{General Relativity and Gravitation}

\begin{document}

\title{Extended bodies with quadrupole moment interacting with gravitational monopoles: reciprocity relations}

\author{Donato Bini \and
        Christian Cherubini \and
        Simonetta Filippi  \and
        Andrea Geralico
}

\institute{Donato Bini 
              \at
              Istituto per le Applicazioni del Calcolo ``M. Picone,'' CNR, I-00161 Rome, Italy\\
              ICRA, University of Rome ``La Sapienza,'' I--00185 Rome, Italy\\
              INFN - Sezione di Firenze, Polo Scientifico, Via Sansone 1, I--50019, Sesto Fiorentino (FI), Italy\\
              \email{binid@icra.it} 
    \and
              Christian Cherubini 
              \at
              Nonlinear Physics and Mathematical Modeling Lab, Engineering Faculty, University Campus Bio-Medico, I-00128 Rome, Italy\\
              ICRA, University of Rome ``La Sapienza,'' I--00185 Rome, Italy\\
              \email{cherubini@icra.it}         
        \and
              Simonetta Filippi 
              \at
              Nonlinear Physics and Mathematical Modeling Lab, Engineering Faculty, University Campus Bio-Medico, I-00128 Rome, Italy\\
              ICRA, University of Rome ``La Sapienza,'' I--00185 Rome, Italy\\
              \email{s.filippi@unicampus.it} 
\and
          Andrea Geralico 
              \at
              Physics Department and ICRA, University of Rome ``La Sapienza,'' I--00185 Rome, Italy\\
              \email{geralico@icra.it}   
}

\date{Received: date / Accepted: date / Version: date}

\maketitle

\begin{abstract}
An exact solution of Einstein's equations representing the static gravitational field of a quasi-spherical source endowed with both mass and mass quadrupole moment is considered. 
It belongs to the Weyl class of solutions and reduces to the Schwarzschild solution when the quadrupole moment vanishes. 
The geometric properties of timelike circular orbits (including geodesics) in this spacetime are investigated.
Moreover, a comparison between geodesic motion in the spacetime of a quasi-spherical source and non-geodesic motion of an extended body also endowed with both mass and mass quadrupole moment as described by Dixon's model in the gravitational field of a Schwarzschild black hole is discussed. 
Certain \lq\lq reciprocity relations" between the source and the particle parameters are obtained, providing a further argument in favor of the acceptability of Dixon's model for extended bodies in general relativity.
\end{abstract}

\section{Introduction}

The fully relativistic multipole moments of a stationary spacetime have been introduced by Hansen \cite{hans}, generalizing previous results by Geroch \cite{ger} valid for static spacetimes only.  Hansen's formulation reduces to Geroch's one in the static limit, in the sense that the recursive definitions of moments are the same but with a different potential.
Beig \cite{Beig1} contributed to clarify Hansen's approach considering a different definition of center of mass; in this way the expansion of the Hansen moments around the center of mass determines the multipole moments uniquely. 
Beig and Simon \cite{Beig2,Simon} also applied the above mentioned definition to the case of stationary axisymmetric spacetimes.
Thorne \cite{Thorne}, before the works of Beig and collaborators, gave another definition of multipole moments which later Gursel \cite{gursel} has shown to be equivalent to Hansen's definition for a source with nonzero rest mass. 
Many exact solutions of Einstein's equations for sources having multipolar structure as described by the Geroch-Hansen formulation are known, mostly belonging to the Weyl class of stationary axisymmetric spacetimes \cite{ES} and 
obtained with a suitable use of Ernst potentials and generating techniques \cite{Ernst}, and hence formally very complicated.

On the other hand, the motion of extended bodies (considered as test bodies, i.e. with backreaction neglected) in any given background was well established after the works of Mathisson, Papapetrou (up to the dipolar structure) \cite{math37,papa51} and Dixon (including any multipolar structure) \cite{dixon64,dixon69,dixon70,dixon73,dixon74}.

In this paper we analyze circular orbits (including geodesics) in the gravitational field of a body endowed with both mass and mass quadrupole moment, which reduces to the familiar Schwarzschild solution when the quadrupole moment vanishes.
Then we compare geodesic motion of a test particle on the equatorial plane with that of an extended body also  endowed with both mass and mass quadrupole moment as described by Dixon's model in the gravitational field of a Schwarzschild black hole. 
We investigate the correspondence between the source and the particle parameters. We obtain certain \lq\lq reciprocity relations" leading to the identification of Dixon's model quadrupole parameters with those underlying Geroch-Hansen approach. 
This is a novel result which should be regarded as a further argument in favour of the acceptability of Dixon's model for extended bodies in general relativity.

\section{Test particles in the field of a quasi-spherical source}

The metric of a nonrotating mass with a quadrupole moment has been obtained long ago by Erez and Rosen \cite{erez}, later corrected for several numerical coefficients independently by Doroshkevich et al. \cite{novikov} and Young and Coulter \cite{young}. 
It is a solution of the static Weyl class of solution with the metric element of the following form
\beq
\label{metric_Weyl}
\rmd s^2 = -e^{2\psi} \rmd t^2 +e^{2(\gamma -\psi)} (\rmd \rho^2 + \rmd z^2)+\rho^2 e^{-2\psi} \rmd z^2 \,,
\eeq
with $\psi$ and $\gamma$ functions of $\rho$ and $z$ only. The associated vacuum Einstein's equations are
\begin{eqnarray}
&& \psi_{\rho\rho}+\frac1\rho \psi_\rho + \psi_{zz}=0\,, \nonumber \\
&& \gamma_\rho= \rho (\psi_\rho^2-\psi_z^2)\,,\quad
\gamma_z= 2\rho \psi_\rho\psi_z \,.
\end{eqnarray}
Within this class, the solution representing a source with mass (monopole) and quadrupole structure can be written in the following form by using prolate spheroidal coordinates $x$ and $y$
\begin{eqnarray}
\psi&=&\psi_0 +q \frac12 (3y^2-1)\left[\frac12 (3x^2-1)\psi_0 +\frac32 x  \right]\,, \\
\gamma&=& \gamma_0+q\left[2\gamma_0-3(1-y^2)\left(x\psi_0+1\right)\right]\,\nonumber \\
&& +q^2 \left\{\gamma_0 +\frac{3}{16}(1-y^2)\left[3(x^2-1)^2\psi_0^2 +2x(3x^2-5)\psi_0+(3x^2-4)\right]  \right. \nonumber \\
&&\left. -\frac{9}{16} y^2(1-y^2)
\left[ (1-x^2)^2(1-9x^2)^2\psi_0^2 +2x(9x^2-7)\psi_0+(3x^2-4)\right]
\right\},\nonumber
\end{eqnarray}
where
\beq
\label{schw}
\psi_0=\frac12 \ln \left(\frac{x-1}{x+1} \right)\,,\quad
\gamma_0=\frac12 \ln \left( \frac{x^2-1}{x^2-y^2}\right)
\eeq
are the monopole quantities corresponding to the Schwarzschild solution (which is obtained for $q=0$). 
The relation with Weyl coordinates $\rho$ and $z$ is given by
\beq
x=\frac{1}{2M}(r_++r_-)\,,\quad y=\frac{1}{2M}(r_+-r_-)\ ,
\eeq
with
\beq
r_\pm=[\rho^2+(z\pm M)^2]^{1/2}\ .
\eeq
Note that on the symmetry hyperplane $z=0$  we have $r_+=r_-=[\rho^2+ M^2]^{1/2}\equiv r_*$ and hence $x=r_*/M$, $y=0$.

It is worth to mention that there exist in the literature several different solutions corresponding to a quasi-spherical source, all with the correct gravitational potential of a massive static source with a quadrupole moment in the Newtonian limit \cite{gutsunaev,hernandez}.
For these metrics some properties of geodesic motion have been analyzed \cite{herrera1,herrera2}. 
Geodesics in the Erez-Rosen spacetime have been also extensively investigated \cite{armenti,quevedo}.
A generalization to the stationary case including rotation of the source has been considered by Quevedo and Mashhoon \cite{quevmash}. 

For our study we refer to the solution given by Young and Coulter \cite{young}.
In the limit of weak fields and small quadrupole moments the nonvanishing Geroch-Hansen moments associated with this solution are the monopole ${\mathcal M}_0=M$ and the quadrupole moment ${\mathcal M}_2=Q$, the latter being related to the parameter $q$ by the equation
\beq
\label{quad_mom}
\frac{Q}{M^3}=\frac2{15}q
\eeq
according to our conventions $G=1=c$.

Due to the stationarity of this spacetime a suitable family of fiducial observers is that of the so called static observers, with unit timelike four velocity $e_{\hat t}= e^{-\psi}\partial_t\,$ aligned with the timelike Killing vector $\partial_t$.
An orthonormal frame adapted to the static observers is given by
\beq
e_{\hat t}= e^{-\psi}\partial_t\,,\quad
e_{\hat \rho}= e^{\psi-\gamma}\partial_\rho\,,\quad
e_{\hat \phi}= \frac{1}{\rho}e^{\psi}\partial_\phi\,,\quad
e_{\hat z}= e^{\psi-\gamma}\partial_z\,.
\eeq

\subsection{Circular orbits}

Circular  orbits ($\rho=\,$const) on a $z=\,$const hypersurface have a four velocity of the form
\beq
\label{Ucirc}
U=\gamma_U (e_{\hat t}+\nu e_{\hat \phi})\,,\quad \gamma_U=(1-\nu^2)^{-1/2}\,,
\eeq
with $\nu$ Lie-constant along $U$, i.e. $\pounds_U \nu=0$.
They form a $1$-parameter family of orbits parametrized by $\nu$.
In general these orbits are accelerated  with a transverse acceleration $a(U)={DU}/{d\tau_U}=a^{\hat \rho}e_{\hat \rho}+a^{\hat z}e_{\hat z}$ such that
\begin{eqnarray}
a_{\hat \rho}=\gamma_U^2 e^{\psi-\gamma}\left[(1+\nu^2)\psi_\rho-\frac{\nu^2}{\rho}  \right]\,,\qquad
a_{\hat z}=\gamma_U^2 e^{\psi-\gamma}(1+\nu^2)\psi_z\,.
\end{eqnarray}
It is convenient to introduce a polar representation for the acceleration components, i.e.
\beq
a_{\hat \rho}=\kappa \cos \chi\,,\qquad a_{\hat z}=\kappa \sin \chi\ ,
\eeq
with
\beq
\kappa=\sqrt{a_{\hat \rho}^2+a_{\hat z}^2}\,,\qquad \tan \chi=a_{\hat z}/a_{\hat \rho}\ .
\eeq
Therefore we have
\beq
a(U)=\kappa (\cos \chi e_{\hat \rho}+\sin \chi e_{\hat z})\equiv \kappa E_1\ .
\eeq
Starting a Frenet-Serret procedure with $U=E_0$  one obtains a Frenet-Serret frame governed by
the transport equations:
\begin{eqnarray}
\label{FSeqs}
\frac{DE_0}{d\tau_U}&=\kappa E_1\ ,\phantom{+\tau_1 E_2\ \ \ } \qquad &  
\frac{DE_1}{d\tau_U}=\kappa E_0+\tau_1 E_2\ ,\nonumber \\
 \nonumber \\
\frac{DE_2}{d\tau_U}&=-\tau_1E_1+\tau_2E_3\ , \qquad &
\frac{DE_3}{d\tau_U}=-\tau_2E_2\ .
\end{eqnarray} 
The curvature $\kappa$ is the magnitude $||a(U)||$ of
the acceleration $a(U)$, while
the first
and second torsions $\tau_1$ and $\tau_2$ are the components of the Frenet-Serret angular velocity vector
\beq
\label{omegaFS}
\omega_{\rm (FS)}=\tau_1 E_3 + \tau_2 E_1\ , \qquad ||\omega_{\rm (FS)}||=[\tau_1^2 + \tau_2^2]^{1/2}\ ,
\eeq
putting the spatial transport equations (\ref{FSeqs}) in the form
\beq\label{eq:DXdtau}
\frac{DE_a}{d\tau_U}=\omega_{\rm (FS)}\times E_a+\kappa E_0\,\delta^1_a\ ,
\eeq
where $\times$ denotes ordinary vector product in the Euclidean three space orthogonal to $U$.
It is easy to show that in this case the Frenet-Serret frame vectors are
\begin{eqnarray}
E_1&=& \cos \chi e_{\hat \rho}+\sin \chi e_{\hat z}\,,\nonumber \\
E_2&=& \gamma_U (\nu e_{\hat t}+ e_{\hat \phi})=\frac{1}{\gamma_U^2}\frac{dU}{d\nu}\,,\nonumber \\
E_3&=& -\sin \chi e_{\hat \rho}+\cos \chi e_{\hat z}=\frac{dE_1}{d\chi}\,,
\end{eqnarray}
while the  Frenet-Serret torsions \cite{circfs,bjdf} are given by
\beq
\tau_1=-\frac{1}{2\gamma_U^2}\frac{d\kappa}{d\nu} \,,\qquad \tau_2=-\frac{\kappa}{2\gamma_U^2}\frac{d\chi}{d\nu}\,  .
\eeq
Moreover, apart from those circular orbits located on the symmetry hyperplane $z=0$ (where $\psi_z=0$), no geodesics exist for whatever special choices of $\nu$. Vice versa, circular orbits on the symmetry plane $z=0$ correspond to 
\beq
\label{kappadef}
\kappa=\gamma_U^2 e^{\psi-\gamma}\frac{1}{\rho}\left[\rho \psi_\rho -\nu^2 (1-\rho \psi_\rho)\right]
\,,\quad
\chi=0
\eeq
implying also 
\beq
\label{tau1def}
\tau_1=\gamma_U^2 e^{\psi-\gamma} \frac{\nu}{\rho}(1-2\rho \psi_\rho)\
\eeq
and $\tau_2=0$.
Note that in this case (symmetry hyperplane) we have fixed $E_1=e_{\hat \rho}$ and allowed $\kappa$ to vary its sign, as it is customary \cite{bjdf}.

A direct evaluation shows that circular geodesics on $z=0$ correspond to
\beq
\label{nugeo}
\nu_{\rm (geo)}^{\pm}=\pm \left[\frac{\rho \psi_\rho}{1-\rho\psi_\rho}\right]^{1/2}\bigg|_{z=0}\ .
\eeq
Using the solution (\ref{metric_Weyl})--(\ref{schw}) we find that 
\beq
\label{nugeoMQ}
\nu_{\rm (geo)}^{\pm}=\pm \left[\frac{q-q_0}{q_1-q}\right]^{1/2}\ ,
\eeq
where the quantities $q_0$ and $q_1$ (which are functions of $r_*/M$) are the quadrupole critical values corresponding to $\rho \psi_\rho=0$ and $\rho \psi_\rho=1$ respectively:
\begin{eqnarray}
\label{q01def}
q_0&=&\left[\frac34 \frac{r_*}{M}\left(\frac{r_*^2}{M^2}-1\right)\ln \left(\frac{r_*-M}{r_*+M} \right)+\frac32\frac{r_*^2}{M^2}-1\right]^{-1}\ , \nonumber\\
q_1&=&q_0\left(1-\frac{r_*}{M} \right)\ .
\end{eqnarray}
Note that at a fixed value of $\rho$ the quantity $\rho\psi_\rho$ is a linear function of $q$.

\begin{figure} 
\typeout{*** EPS figure 1ab}
\begin{center}
$\begin{array}{cc}
\includegraphics[scale=0.4]{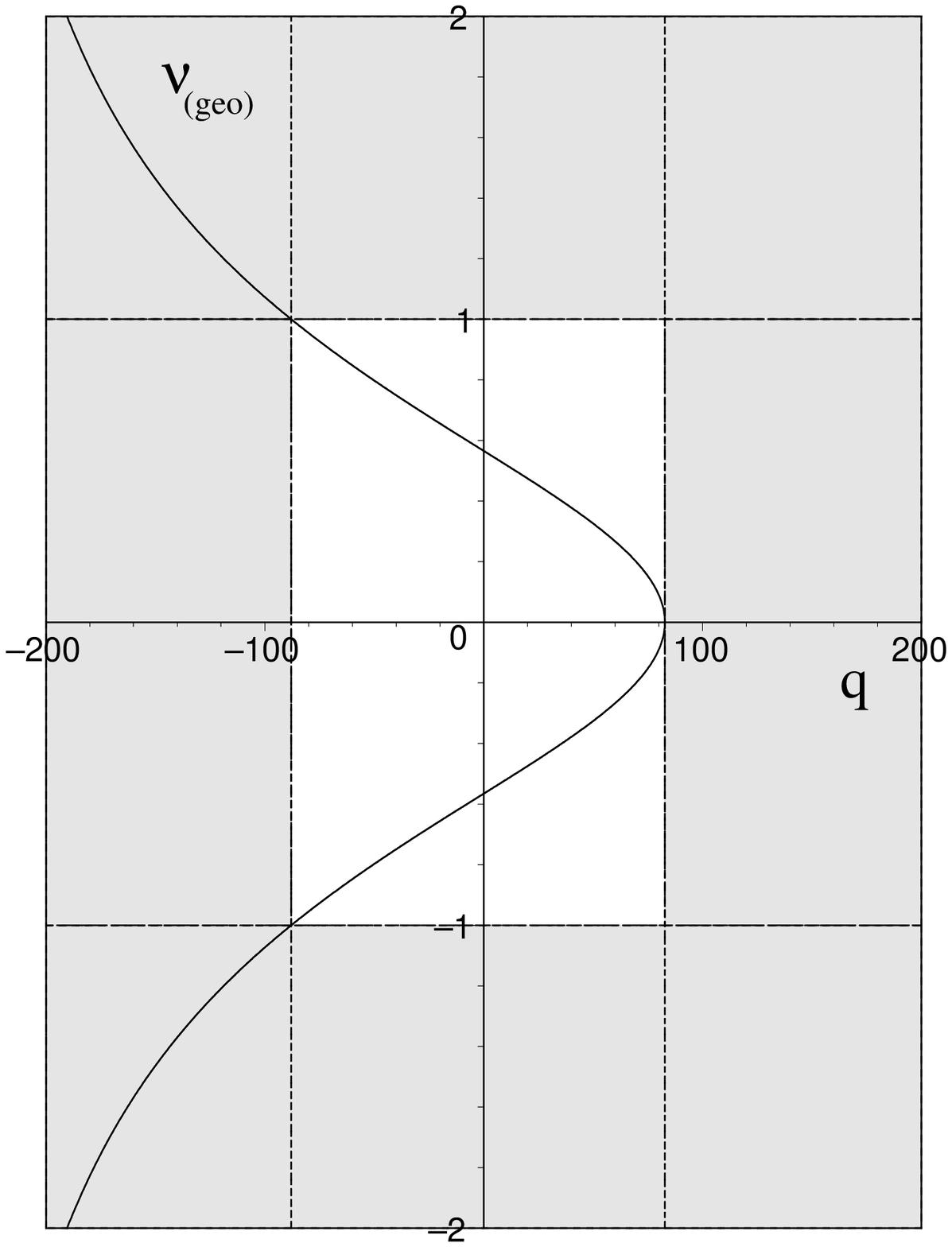}&\qquad
\includegraphics[scale=0.4]{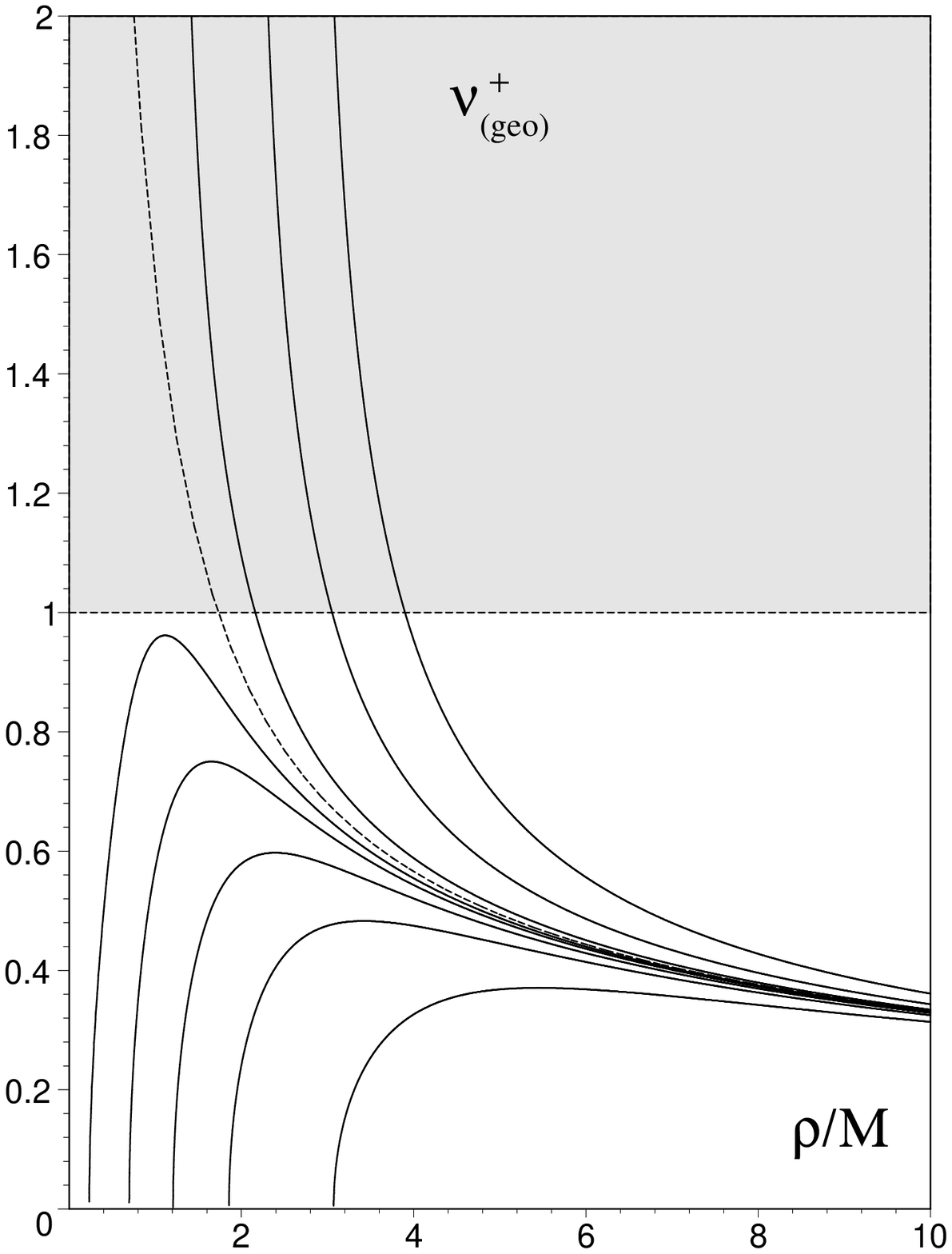}\\[0.4cm]
\mbox{(a)} & \mbox{(b)}\\
\end{array}$
\end{center}
\caption{The geodesic linear velocities $\nu_{\rm (geo)}^{\pm}$ are plotted in Fig. (a) as functions of the quadrupole parameter $q$ for fixed radial distance $\rho/M=4$ from the source. 
Timelike circular geodesics exist for $q_2<q<q_0$ (see Eq. (\ref{rangefin})), with $q_0\approx82.84$ and $q_2\approx-87.94$ for this choice of parameters.
The behaviour of $\nu_{\rm (geo)}^{+}$ as a function of $\rho/M$ is shown in Fig. (b) for different values of $q=[-80,-30,-5,0,2.5,5,10,20,50]$.
The case $q=0$ (Schwarzschild) corresponds to the dashed curve without a relative maximum.
The shape of the curves with negative $q$ is very similar to the Schwarzschild one, the lightlike condition $\nu_{\rm (geo)}^{+}=1$ being reached at greater values of the radial distance for decreasing values of $q$.
The curves with positive $q$ instead all present a relative maximum and are ordered from left to right for increasing values of $q$.
Shaded regions are forbidden.
}
\label{fig:1}
\end{figure}

The geodesic velocities (\ref{nugeoMQ}) are plotted in Fig. \ref{fig:1} both as functions of the quadrupole parameter $q$ for fixed radial distance (see Fig. (a)) and as functions of $\rho/M$ for different values of $q$ (see Fig. (b)).
In the first case (Fig. (a)) we have shown how the quadrupole moment affects the causality condition: there exist a finite range of values of $q$ wherein timelike circular geodesics are allowed (see the discussion below).
The difference from the Schwarzschild case is clear instead from Fig. (b): the behaviour of the velocities differs significantly at small distances from the source, whereas it is quite similar for large distances.


For a later use it is useful to consider the following two limits of $\nu_{\rm (geo)}^{\pm}$: 
$M/\rho \to 0$ for fixed $q$ and $q\to 0$ for fixed $M$ and $\rho$.
In the first case ($M/\rho \to 0$) we have
\beq
\label{nugeo_asympt}
\nu_{\rm (geo)}^{\pm} = \pm \left(\frac{M}{\rho} \right)^{1/2}
\left[  
1+\frac12 \left(\frac{M}{\rho}\right)-\frac14 \left(\frac{M}{\rho}\right)^2 \frac{4q-5}{10}
\right]+O\left(\frac{M}{\rho}\right)^{7/2}\,.
\eeq
In the second case ($q\to 0$) we have instead
\beq
\label{eq_da_cfr}
\nu_{\rm (geo)}^{\pm}=\pm\nu_K \pm q W(\nu_K)+O(q^2)\ ,
\eeq
where
\beq
W(\nu_K)=\frac{1+\nu_K^2}{8\nu_K^5}\left[3(1+\nu_K^2)(1+2\nu_K^2)\ln (1+2\nu_K^2)-2\nu_K^2(\nu_K^4+6\nu_K^2+3)\right]\ 
\eeq
is a function of the linear velocity $\nu_K$ of circular geodesics in the Schwarzschild spacetime given by
\beq
\label{nuKdef}
\nu_K=\sqrt{\frac{M}{r_*-M}}\ .
\eeq
It is easy to show that in the limit $M/\rho \to 0$ one has
\begin{eqnarray}
\nu_K &\to& \left(\frac{M}{\rho} \right)^{1/2} \left[1+\frac12 \left(\frac{M}{\rho} \right)+\frac18 \left(\frac{M}{\rho} \right)^2 \right]+ O\left(\frac{M}{\rho}\right)^{7/2}\ , \nonumber \\ 
W(\nu_K)&\to&  -\frac{1}{10}\left(\frac{M}{\rho}\right)^{5/2} + O\left(\frac{M}{\rho}\right)^{7/2}\ .
\end{eqnarray}

\begin{figure}
\typeout{*** EPS figure 2}
\begin{center}
\includegraphics[scale=0.4]{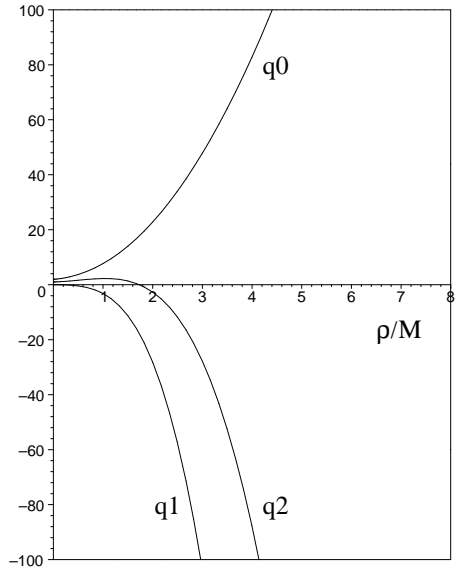}
\end{center}
\caption{The behaviours of the critical quadrupole parameters $q_0$, $q_1$ and $q_2$ are shown as functions of $\rho/M$.
}
\label{fig:2}
\end{figure}

Let us discuss now the causal properties of geodesics.
The requirement that the argument of the square root of Eq. (\ref{nugeo}) be positive gives
\beq
\frac{\rho \psi_\rho}{1-\rho\psi_\rho}>0 
\qquad \to \qquad 0<\rho \psi_\rho<1
\qquad \to \qquad q_1<q<q_0\ ,
\eeq
where the quantities $q_0$ and $q_1$ are given by Eq. (\ref{q01def}).
The further requirement that $|\nu_{\rm (geo)}^{\pm}|<1$ ensures that they remain timelike.
This condition implies  
\beq
\label{q2def}
q>\frac12(q_0+q_1)\equiv q_2\ ,
\eeq
so that the condition for the existence of timelike circular geodesics turns out to be
\beq
\label{rangefin}
q_2<q<q_0\ .
\eeq
Fig. \ref{fig:2} shows the behaviours of the critical quadrupole parameters $q_0$, $q_1$ and $q_2$ as functions of the radial distance $\rho/M$.


\begin{figure} 
\typeout{*** EPS figure 3ab}
\begin{center}
$\begin{array}{cc}
\includegraphics[scale=0.4]{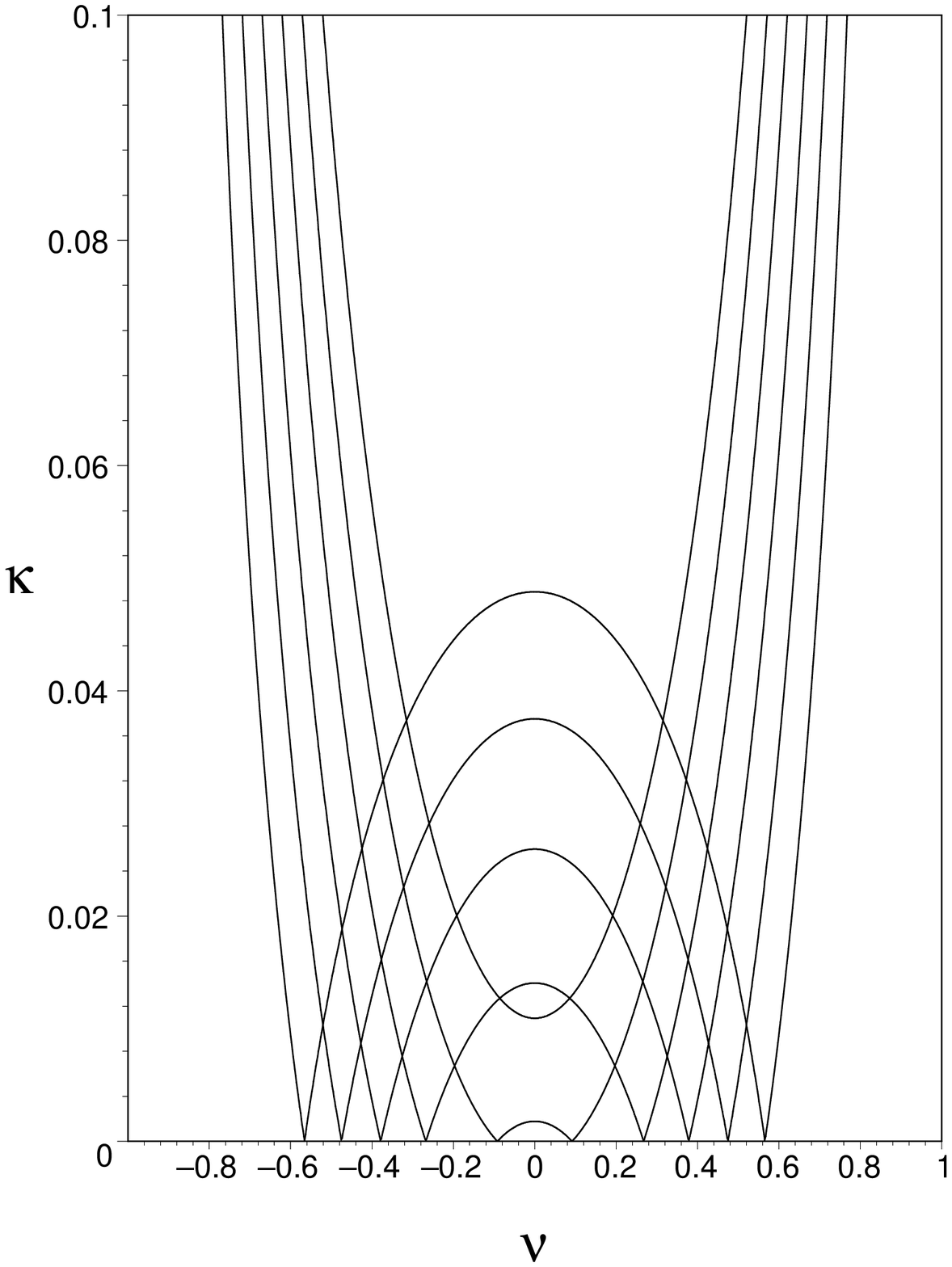}&\qquad
\includegraphics[scale=0.4]{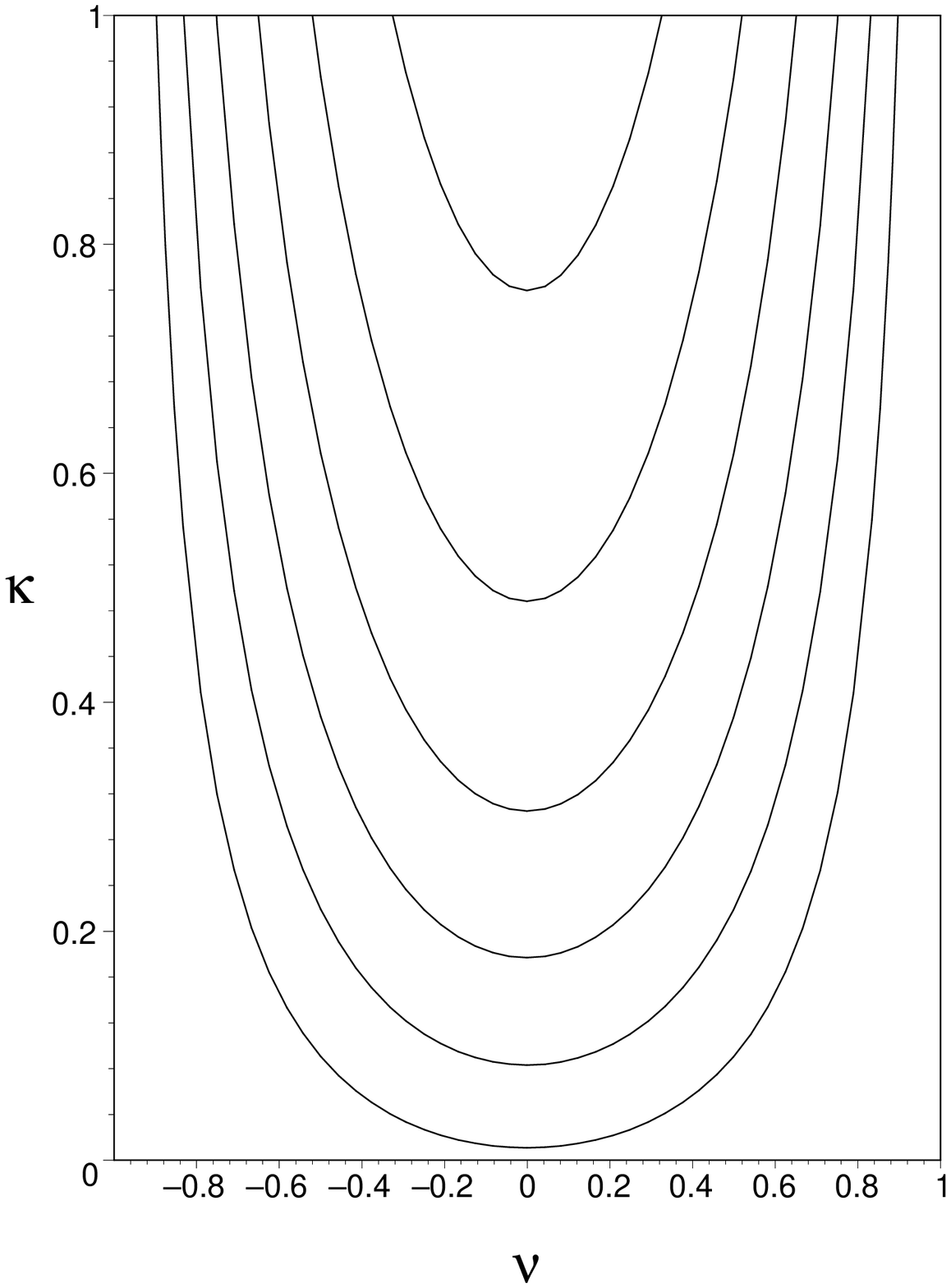}\\[0.4cm]
\mbox{(a)} & \mbox{(b)}\\
\end{array}$
\end{center}
\caption{The magnitude of the acceleration $\kappa$ for circular orbits at $z=0$ is plotted as a function of $\nu$ for $\rho/M=4$ and different values of the quadrupole parameter: (a) $q=[0,20,40,60,80,100]$ and (b) $q=[100,200,300,400,500,600]$. 
The (symmetric) values of $\nu$ associated with $\kappa=0$ correspond to geodesics, i.e. $\nu_{\rm (geo)}^{\pm}$ (see Fig. (a)). 
The various curves are ordered so that for increasing values of $q$ the value of $|\nu_{\rm (geo)}^{\pm}|$ decreases until it approaches the critical value $q_0\approx82.84$ beyond which circular geodesics do not exist anymore (see Fig. (b)).
It is worth noting that the situation is very similar for selected negative values of $q$. 
}
\label{fig:3}
\end{figure}

In terms of the quantities $q_0$ and $q_1$ the expression (\ref{kappadef}) for the acceleration $\kappa$ becomes
\beq
\kappa=\gamma_U^2 e^{\psi-\gamma}\frac{1}{\rho}\left(\frac{q-q_1}{q_1-q_0}\right)\left[\nu^2-(\nu_{\rm (geo)}^{\pm})^2\right]\ .
\eeq
Its behaviour as a function of $\nu$ is shown in Fig. \ref{fig:3} for different values of the quadrupole parameter. 
For increasing values of $q$ the corresponding values of $\nu_{\rm (geo)}^{\pm}$ decrease up to a critical value (corresponding to $\nu_{\rm (geo)}^{\pm}=0$) beyond which they do not exist anymore.


\begin{figure} 
\typeout{*** EPS figure 4ab}
\begin{center}
$\begin{array}{cc}
\includegraphics[scale=0.4]{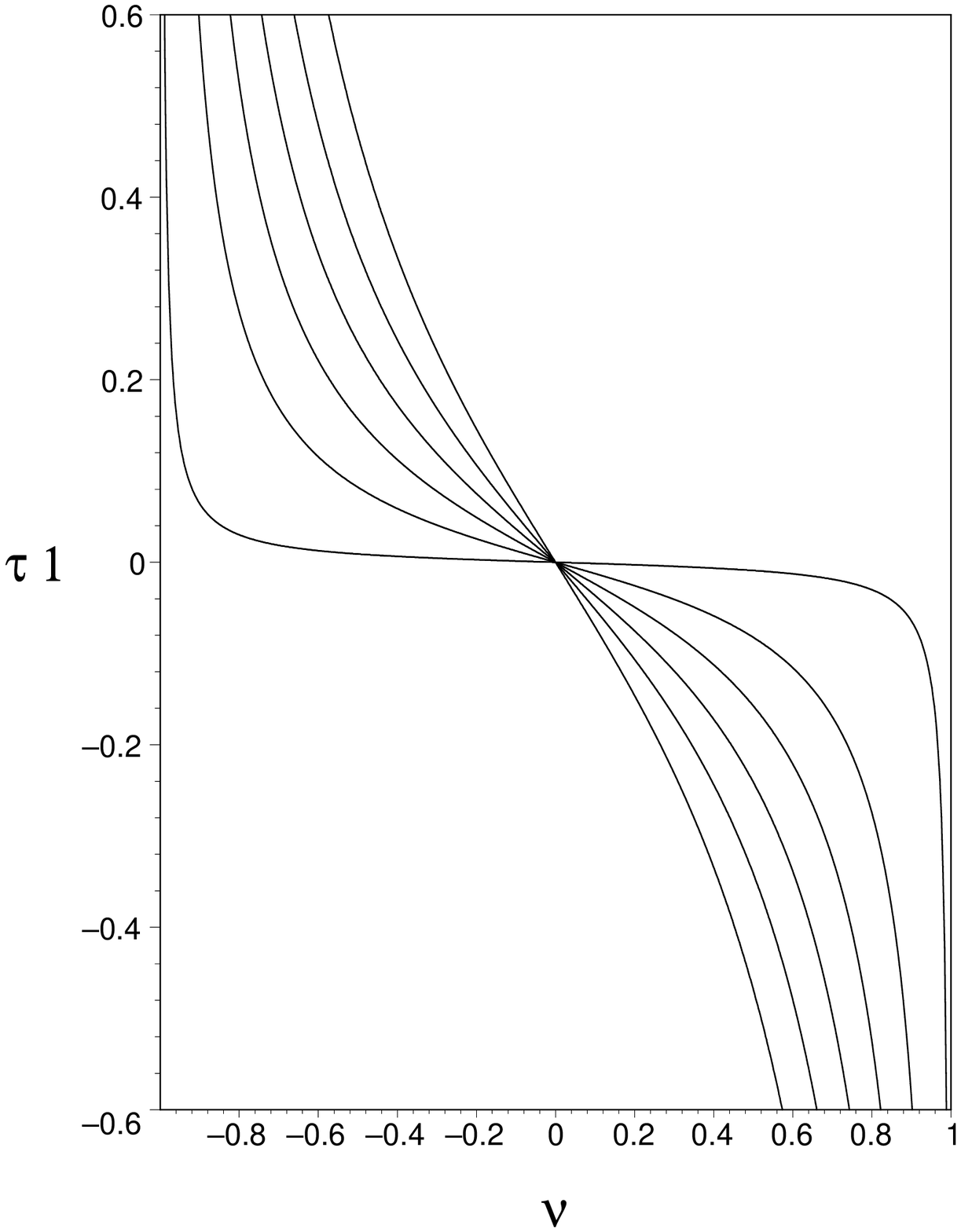}&\qquad
\includegraphics[scale=0.4]{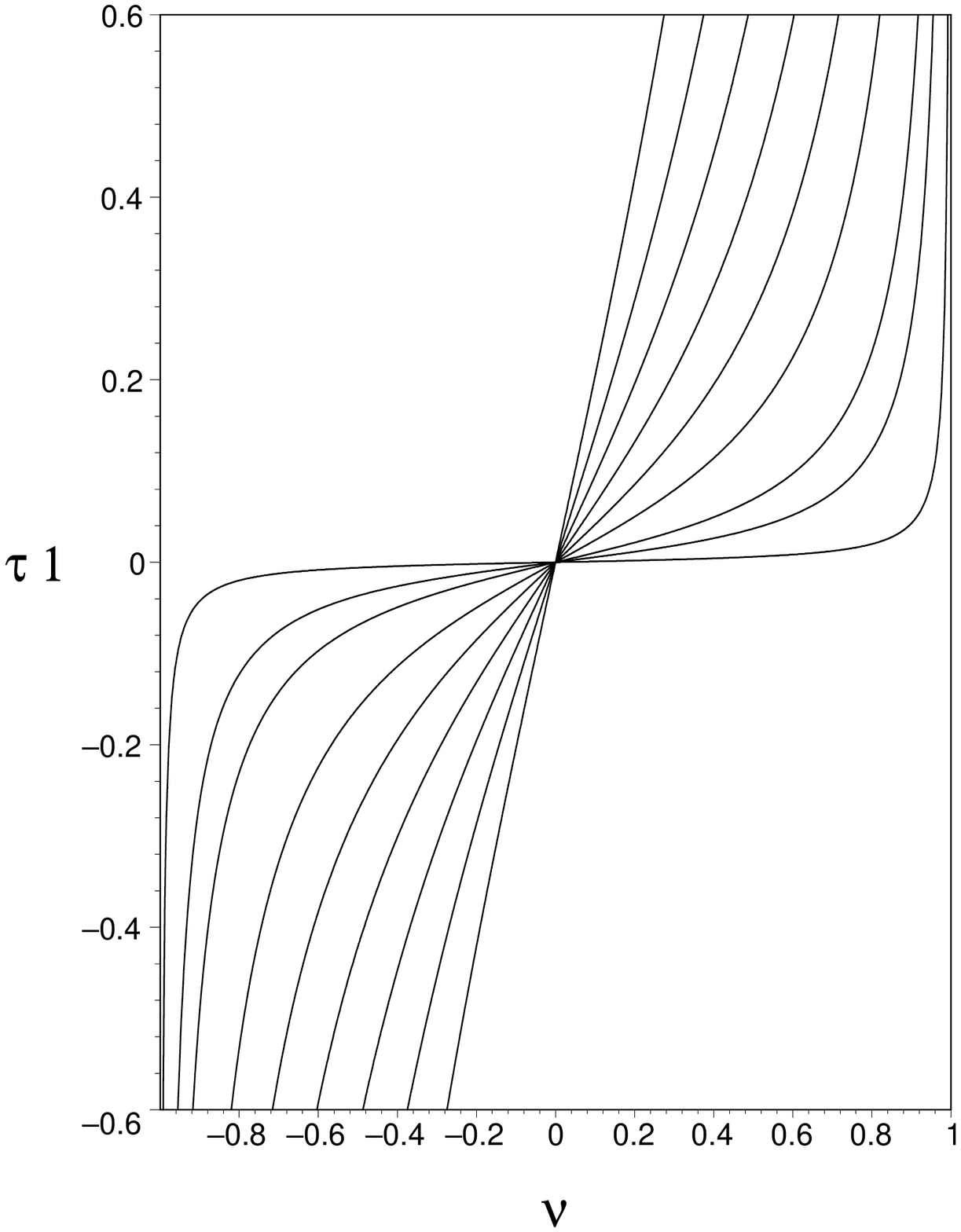}\\[0.4cm]
\mbox{(a)} & \mbox{(b)}\\
\end{array}$
\end{center}
\caption{The first torsion $\tau_1$ for circular orbits at $z=0$ is plotted as a function of $\nu$ for $\rho/M=4$ and different values of the quadrupole parameter: (a) $q=[-600,-500,-400,-300,-200,-100]$ and (b) $q=[-80,-40,0,100,200,300,400,500,600]$.
The behaviour changes depending on whether $q$ is less or greater than the critical value $q_2\approx-87.94$.
The curves in both Figs. (a) and (b) are ordered in such way that for increasing $|q|$ they shrink on the vertical axis. 
}
\label{fig:4}
\end{figure}

A similar discussion can be done for the first torsion, i.e. there exists a critical value of the quadrupole moment $q$ such that $\tau_1=0$ for any $\nu$. This happens when $\rho \psi_\rho=1/2$. 
It is quite surprising that the corresponding critical value of $q$ coincides with the limiting value $q_2$ for timelike geodesics as defined in Eq. (\ref{q2def}).
In fact, by introducing the quantities $q_0$, $q_1$ and $q_2$ Eq. (\ref{tau1def}) can be cast in the form
\beq
\tau_1=\gamma_U^2 e^{\psi-\gamma} \frac{\nu}{\rho}\left(\frac{q_2-q}{q_1-q_0}\right)\ .
\eeq
We thus recognize how the presence of the quadrupole moment $q$ changes completely and enriches the situation with respect to the Schwarzschild case, examined long ago by many authors \cite{circfs,bjdf}.
This is well elucidated in Fig. \ref{fig:4}, where the behaviour of $\tau_1$ as a function of $\nu$ is discussed for different values of the quadrupole parameter.


\section{Motion of quadrupolar particles in the Schwarzschild background}

We consider now the complementary case of an extended body with structure up to the quadrupole in the field of a Schwarzschild background, following the description due to Dixon \cite{dixon64,dixon69,dixon70,dixon73,dixon74}.
In the quadrupole approximation Dixon's equations are\footnote{
Note that the torque term $D^{\rm (quad)}{}^{\mu\nu}$ in the second set of equations was misprinted in  previous works \cite{quad1,quad2,quad3,quad4}.
The results of all those papers are but correct. 
}
\begin{eqnarray}
\label{papcoreqs1}
\frac{DP^{\mu}}{\rmd \tau_U}&=&-\frac12R^{\mu}{}_{\nu\alpha\beta}U^{\nu}S^{\alpha\beta}-\frac16J^{\alpha\beta\gamma\delta}R_{\alpha\beta\gamma\delta}{}^{;\,\mu}
\equiv F^{\rm (spin)}{}^{\mu}+F^{\rm (quad)}{}^{\mu}\ , \\
\label{papcoreqs2}
\frac{DS^{\mu\nu}}{\rmd \tau_U}&=&2P^{[\mu}U^{\nu]}+\frac43J^{\alpha\beta\gamma[\mu}R^{\nu]}{}_{\gamma\alpha\beta}
\equiv2P^{[\mu}U^{\nu]}+D^{\rm (quad)}{}^{\mu\nu}\ ,
\end{eqnarray}
where $P^{\mu}=m U_p^\mu$ (with $U_p\cdot U_p=-1$) is the total four-momentum of the particle, and $S^{\mu\nu}$ is a (antisymmetric) spin tensor; 
$U$ is the timelike unit tangent vector of the \lq\lq center of mass line'' ${\mathcal C}_U$ used to make the multipole reduction, parametrized by the proper time $\tau_U$.
The tensor $J^{\alpha\beta\gamma\delta}$ is the quadrupole moment of the stress-energy tensor of the body, and has the same algebraic symmetries as the Riemann tensor, i.e. 20 independent components. 
Note that there are several equivalent expressions used in the literature for the torque $D^{\rm (quad)}{}^{\mu\nu}$, which can be summarized by 
\beq
D^{\rm (quad)}_{{\rm dixon}}{}^{\mu\nu}=-\frac43R^{[\mu}{}_{\alpha\beta\gamma}J^{\nu]\alpha\beta\gamma}\ , \quad
D^{\rm (quad)}_{{\rm h-r}}{}^{\mu\nu}=\frac43J^{\alpha\beta\gamma[\mu}R^{\nu]}{}_{\gamma\alpha\beta}\ ,
\eeq
due to Dixon (\cite{dixon74}, p. 65) and Ehlers and Rudolph (\cite{ehlers77}, p. 209).

There are no evolution equations for the quadrupole as well as higher multipoles as a consequence of the Dixon's construction, so the structure only depends on the considered body.
Therefore the system of equations is not self-consistent, and one must assume that all unspecified quantities are  known as intrinsic properties of the matter under consideration.

Moreover, in order the model to be mathematically correct certain additional conditions should be imposed \cite{dixon64}. 
As it is standard one may limit considerations to Dixon's model under the further simplifying assumption \cite{taub64,ehlers77} 
that the only contribution to the complete quadrupole moment $J^{\alpha\beta\gamma\delta}$ stems from the mass quadrupole moment $Q^{\alpha\beta}$ so that
\beq
\label{Jdef}
J^{\alpha\beta\gamma\delta}=-3U_p^{[\alpha}Q^{\beta][\gamma}U_p^{\delta]}\ ,\qquad Q^{\alpha\beta}U_p{}_\beta=0\ .
\eeq

We are interested here in bodies with vanishing spinning structure, so that the model equations undergo big simplifications and reduce to
\begin{eqnarray}
\label{papcoreqs1new}
m\frac{D U_p^{\mu}}{\rmd \tau_U}
&=& \frac12 U_p^{\alpha}Q^{\beta\gamma}U_p^{\delta}R_{\alpha\beta\gamma\delta}{}^{;\,\mu}\ , \\
\label{papcoreqs2new}
m U_p^{[\mu}U^{\nu]}&=&U_p^\alpha Q^{\beta\gamma}U_p^{[\mu}R^{\nu]}{}_{\gamma\alpha\beta}-U_p^\alpha U_p^\gamma Q^{\beta [\mu}R^{\nu]}{}_{\gamma\alpha\beta}\ ,
\end{eqnarray}
where we have also assumed for simplicity the mass of the body as a constant.

Let us consider the case of the Schwarzschild spacetime, with the metric given by Eq. (\ref{metric_Weyl}) with
$\gamma=\gamma_0$ and $\psi=\psi_0$,
given by Eq. (\ref{schw}).

Let us assume that $U$ is tangent to a (timelike) spatially circular orbit 
\beq
\label{orbita}
U=\gamma_U [e_{\hat t} +\nu e_{\hat \phi}], \qquad \gamma_U=(1-\nu^2)^{-1/2}\ ,
\eeq
with $\nu$ constant along $U$.  
We limit our analysis to  the equatorial plane ($z=0$) of the Schwarzschild solution.
The linear velocity corresponding to circular geodesics is given by Eq. (\ref{nuKdef}), whereas the angular velocity is given by 
\beq
\zeta_K=\sqrt{\frac{M}{(r_*+M)^3}}=\frac{\nu_K^3}{M(1+2\nu_K^2)^{3/2}}\ .
\eeq

Let also $P=m\, U_p$ be such that
\begin{equation}
\label{Ptot}
U_p=\gamma_p\, [e_{\hat t}+\nu_p e_{\hat \phi}]\ , \quad \gamma_p=(1-\nu_p^2)^{-1/2}\ , 
\end{equation}
i.e. let us assume that $U_p$ also is tangent to a circular orbit and
set up an orthonormal frame adapted to $U_p$ given by
\beq
e_0=U_p\ , \qquad e_1=e_{\hat \rho}\ , \qquad e_2=e_{\hat z}\ , \qquad e_3=\gamma_p\, [\nu_p e_{\hat t}+ e_{\hat \phi}]\ ;
\eeq
hereafter all frame components of the various fields  are  meant to be referred to such a frame.
Note that the assumption of having both $U$ and $U_p$ leads to great simplifications to Dixon's equations.

From Eq. (\ref{Jdef})$_2$ we have 
\beq
Q_{00}=Q_{01}=Q_{02}=Q_{03}=0\ .
\eeq
We also assume that that all the surviving components of the mass quadrupole moment are all constant along the path.  The latter  assumption corresponds to the definition of \lq\lq quasirigid motion" (or \lq\lq quasirigid bodies") due to Ehlers and Rudolph \cite{ehlers77}.  
Clearly in a more realistic situation the latter hypothesis should be released.

Consider first the constraint equations (\ref{papcoreqs2new}). They imply that 
\beq
Q_{12}=Q_{13}=Q_{23}=0\ ,
\eeq
and after introducing the following \lq\lq structure functions" of the extended body
\beq
Q_{11}=Q_{22}-f, \qquad Q_{33}=Q_{22}-f'\ , 
\eeq
they also give
\beq
\label{moto1}
0=\gamma_U(\nu-\nu_p)-3\gamma_p\nu_p\frac{\zeta_K^2}{m}f\ .
\eeq
Note that the quantities $f$ and $f'$ are necessarily small, in order to avoid backreaction effects.

Consider then the equations of motion (\ref{papcoreqs1new}). 
They imply that
\beq
\label{moto2}
0=\gamma_U\gamma_p (\nu\nu_p-\nu_K^2)-\frac32\frac{\zeta_K^2}{m}[f'+(1-3\gamma_p^2)f]\ .
\eeq
Solving Eqs. (\ref{moto1}) and (\ref{moto2}) for $\nu$ and $\nu_p$ in terms of $f$, $f'$ completely determines the motion.

The quadrupolar structure of the body turns out to be represented by three quantities: $f$, $f'$ and $Q_{22}$. However,
classically, the  quadrupole moment tensor of a mass distribution is tracefree. Assuming for simplicity the same property to hold also for the relativistic quadrupole moment tensor implies
\beq
0=Q_{11}+Q_{22}+Q_{33}=3Q_{22}-f-f'\ ,
\eeq
so that the components  $Q_{ab}$ in this case are completely determined by 
$f$ and $f'$ only
\beq
Q_{11}=-\frac23 f+\frac13 f'\ , \quad Q_{22}=\frac13 (f+f')\ , \quad
Q_{33}=\frac13f-\frac23 f'\ .
\eeq
If the body is axially symmetric about the $z$-axis, then $f'=f$ and the frame components of $Q$ reduce to
\beq
Q_{ab}={\rm diag}\,[-f/3,2f/3,-f/3]\ .
\eeq
It is worth noting that the above assumptions for the extended body's structure (i.e. constant frame components for the quadrupole tensor represented by a single structure function $f$) here used in order to make our analysis as simple as possible can be easily relaxed in favour of a more realistic description.

Eqs. (\ref{moto1}) and (\ref{moto2}) then become
\begin{eqnarray}
\label{moto1_new}
0&=&\gamma_U(\nu-\nu_p)-3\gamma_p\nu_p\zeta_K^2\frac{f}{m}\ , \\
\label{moto2_new}
0&=&\gamma_U\gamma_p (\nu\nu_p-\nu_K^2)-\frac32(2-3\gamma_p^2)\zeta_K^2\frac{f}{m}\ .
\end{eqnarray}
The above relations define the kinematical conditions allowing circular motion of the extended body taking into account its quadrupolar structures. 

Due to the smallness of quadrupolar quantity $f$ the motion of the extended body will be nearly geodesic with correction which we will retain up to first order in $f$.   
Eqs. (\ref{moto1_new}) and (\ref{moto2_new}) thus imply
\beq
\label{nusol}
\nu_\pm\simeq\pm\left[\nu_K-\frac{3\zeta_K^2}{4\nu_K} \frac{f}{m}\right]\ , \qquad 
\nu_p^{(\pm)}\simeq\nu_\pm \mp 3 \nu_K\zeta_K^2\frac{f}{m}\ ,
\eeq
where the signs $\pm$ correspond to co/counter rotating orbits.
The corresponding angular velocity $\zeta_\pm=(\zeta_K/\nu_K)\nu_\pm$ and its reciprocal are
\begin{eqnarray} 
\label{zetasol}
\zeta_\pm \simeq  \pm\zeta_K \left[1 - \frac{3 \zeta_K^2}{4\nu_K^2}\frac{f}{m}\right]\ , \qquad
\frac1{\zeta_\pm} \simeq  \pm\frac{1}{\zeta_K}\left[1 + \frac{3\zeta_K^2}{4\nu_K^2}\frac{f}{m}\right]\ .
\end{eqnarray}

\section{Quadrupolar particles interacting with gravitational monopoles}

The linear velocity (\ref{nusol})$_1$ of the center of mass of the extended body described using Dixon's model, namely
\beq
\label{nusol_rep}
\nu_\pm\simeq\pm\left[\nu_K-\frac{3\zeta_K^2}{4\nu_K} \frac{f}{m}\right]
\eeq
should now be compared with the geodesic linear velocities $\nu_{\rm (geo)}^{\pm}$ given by Eq. (\ref{eq_da_cfr}),
in the limits in which comparison is really allowed (i.e. in the limit of validity of Dixon's model): small mass and quadrupole moment of the body if compared with the mass of the central object. 
Note that a similar treatment was done in the case of a spinning particle orbiting a Schwarzschild black hole compared with a geodesic in the Kerr spacetime \cite{bdfgschwclock}.

The result (to first order in both $q$ and $f$) is
\beq
\label{confr}
-\frac{3\zeta_K^2}{4\nu_K} \frac{f}{m} = q\, W(\nu_K)\,.
\eeq 
In the limit of large distances from the central source ($\rho\gg M$) Eq. (\ref{confr}) gives the nice result
\beq
\frac{f}{m}=\frac2{15}M^2q=\frac{Q}{M}\ .
\eeq
This correspondence validates once more Dixon's model.

\section{Concluding remarks}

We have considered the static gravitational field of a quasi-spherical source belonging to the Weyl class of solutions of Einstein's equations. In this field we have discussed  the geometric properties of (accelerated in general) timelike circular orbits, by analyzing the associated Frenet-Serret curvature and torsions. Among these orbits we have studied in detail the geodesics on the equatorial plane and found a number of interesting properties. For instance we have found that there exists a critical value for the quadrupole moment beyond which the are no more timelike circular geodesics. This fact was unexpected, on the basis of previous works concerning static Schwarzschild black hole spacetime. 

However, the most important contribution of this paper concerns comparison between geodesic motion in the field of a quasi-spherical source with the motion of an extended body (quadrupolar particle, described by using Dixon's model) in the background of a Schwarzschild black hole.
We have shown how the quadrupole moment of a gravitational source as described by the Geroch-Hansen approach coincides with the quadrupole moment of a Dixon's extended body.

\section*{Acknowledgments}

D.B. and A.G. thank P. Fortini and A. Ortolan for their contribution in starting the study of quadrupolar particles by using Dixon's model.
All authors acknowledge ICRANet for support.
The anonimous referees are also thanked for useful comments, especially for bringing to our attention a misprint in the previous version of our Eq. (\ref{papcoreqs2}).

\end{document}